 \documentclass[prl,aps,epsf,superscriptaddress]{revtex4}
 \usepackage{graphicx}
  
\begin{document}
  \title{Giant oscillations of the density of states and the conductance in a  ferromagnetic conductor coupled to
  two superconductors}
  \author{A. Kadigrobov
  }
   \affiliation{
  Theoretische Physik III, Ruhr-Universität Bochum, D-44780 Bochum, Germany
  }
  \affiliation{
  Department of Physics, G\"oteborg University, SE-412 96 G\"oteborg, Sweden
  }
  \author{R. I. Shekhter
  }
  \affiliation{
  Department of Physics, G\"oteborg University, SE-412 96 G\"oteborg, Sweden
  }
  \author{M.~Jonson}
  \affiliation{
  Department of Physics, G\"oteborg University, SE-412 96 G\"oteborg, Sweden
  }

\begin{abstract}
Giant oscillations of the density of electronic states and the differential conductance of a
superconductor-ferromagnet-superconductor structure  are
predicted for the case that the exchange energy of the interaction between the electron spin and
the spontaneous moment of the ferromagnet, $I_0$, is less than the
superconductor energy gap, $\Delta$ ($I_0 <\Delta$ ).
 The effect is due to an extremely large degeneration
of the energy level $\varepsilon = I_0$ ($\varepsilon$ is the electron energy
measured from the Fermi-energy) if the superconductor phase difference $\varphi$ is
close to odd numbers of $\pi$.
 These quantum
interference effect persists even in long ferromagnetic bridges the length
of which much exceeds the ''magnetic length'' $\hbar v_{F}/I_{0}$ for the ballistic case and $\sqrt{\hbar D/I_0}$
for the diffusive one ($D$ is the electron diffusion constant). The predicted effect
allows a direct spectroscopy of Andreev levels in the ferromagnet as well as
a direct measurement of the exchange energy,$I_0$, of
the interaction of the electron spin with
the spontaneous moment of the ferromagnet.
\end{abstract}

  \maketitle
  \section{Introduction}

In recent years much attention has been paid to the conductance of
mesoscopic superconductor-normal conductor- superconductor systems (S/N/S
heterostructures) which is  sensitive to the phase difference between the
superconductors (see, e.g., the review paper by Lambert and Rimondi
\cite{Lamb} and references there). This effect takes place both in ballistic and diffusive samples due
to a quantum interference caused by Andreev reflections of quasi-particles
at two (or more) N/S interfaces which imposes the phase of the
superconducting condensate on the quasiparticle wave function in the normal
conductor.

\begin{figure}
 \centerline{\includegraphics[width=8.0cm]{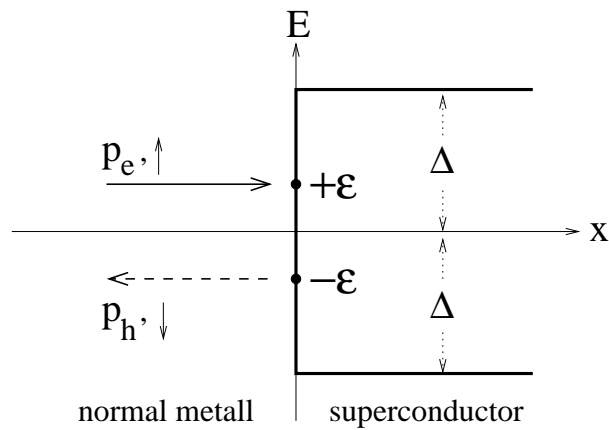}}
  \vspace*{2mm}
  \caption{Schematic representation of the Andreev reflection: the incident electron with
 energy $\varepsilon$ (measured from the Fermi energy $\varepsilon_F$) and the spin up   is reflected back
 as a hole with the energy $-\epsilon$ and the spin down; the incident electron and the reflected hole momenta are
 $p_e=\sqrt{p_F^2 +2m\varepsilon}$ and $p_e=\sqrt{p_F^2 -2m\varepsilon}$
 ($p_F$ is the Fermi momentum, m is the electron mass), respectively.}
\label{Andreev}
  \end{figure}

\begin{figure}
 \centerline{\includegraphics[width=8.0cm]{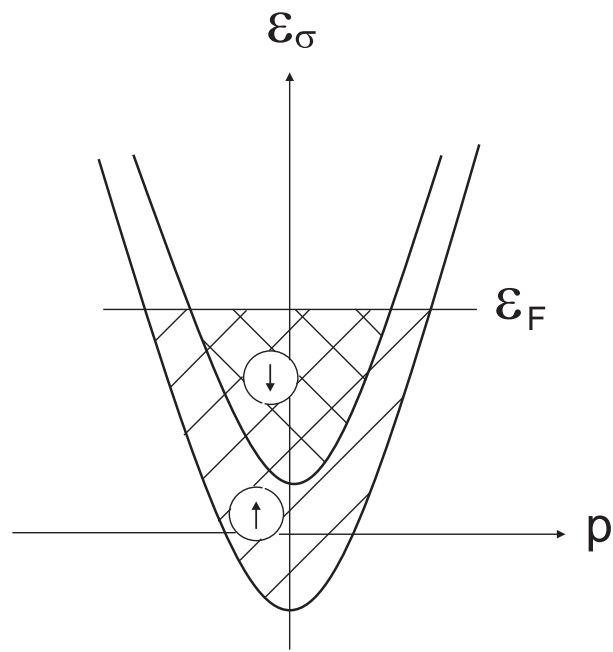}}
  \vspace*{2mm}
  \caption{Energy bands for electrons with opposite spins}
\label{zbband}
  \end{figure}

In normal conductor-superconductor heterostructures, electronic elementary excitations which freely
propagate inside the non-superconducting metal can not penetrate into the
superconductor if their energy $\varepsilon $ (measured from the Fermi
energy) is less than the superconductor energy gap $\Delta $. A correlated
transferring of two electrons accompanied by their pairing  inside the
superconductor is the only mechanism that provides a direct transmission
of the charge into the superconducting condensate that is the ground state
of the superconductor. This 2-electron transmission may be considered as a
scattering process under which electronic excitations inside the normal
conductor  undergo an electron-hole transformation at the boundary with the
superconductor. This scattering (known as the Andreev reflection) couples
the incident electron (hole) and the reflected hole (electron) in such a way that their spins
are oriented in opposite directions and their energies ($
\pm \varepsilon $) are symmetrically positioned with respect to\ the Fermi
energy $\varepsilon _{F}$ (''Andreev hybrid''), as  shown in
Fig.\ref{Andreev}.
As a result of the Andreev reflection the  electronic excitations reflected at the
normal conductor - superconductor interface pick up the phase $\phi$  of the superconductor
order parameter $\Delta = |\Delta|\exp{i \phi}$ and keep memory of  it  inside the
normal conductor.  Such a 2-electron
superconducting correlation persists inside the normal conductor at the
distance ${cal L}_{\epsilon }$ from the superconductor,
\begin{equation}
{\cal L}_{\epsilon
}=min\left(\hbar /\Delta p,\sqrt{\hbar D/\Delta pv_{F}}\right)
\label{corrdistance}
\end{equation}
 where $\Delta
p=p_e -p_h \sim \varepsilon /v_{F}$, $v_{F}$ is the Fermi velocity and $D$ is the
diffusion coefficient. At larger distances from the superconductor  the destruction of
the phase coherence arises due to
the difference between the momenta of the electron ($p_e =p_F +\varepsilon /v_{F}$)
and hole ($p_h=p_F -\varepsilon /v_{F}$) components in the Andreev hybrid (the typical value of the
energy is  $\varepsilon \sim kT$,
where $T$ is the temperature and $k$ is the Boltzmann constant). The other
peculiar feature of the Andreev hybrid is that the electron and hole spins
have opposite directions. This spin flip does not change the interference
pattern of the non-magnetic metal because all energy levels are twice
degenerated with respect to the spin direction. In ferromagnets, however,
this degeneracy is lifted due to the interaction of the electron spin with
the spontaneous moment of the ferromagnet (below we refer to it as the
exchange-interaction energy $I_{0}$), and electrons with opposite directions
of the spins occupy different energy bands (see Fig.\ref{zbband} ):
\begin{equation}
\varepsilon _{\sigma }({\bf p})={\bf p}^{2}/2m+\sigma I_{0}  \label{band}
\end{equation}
where ${\bf p}$ is the electron momentum, $m$ is the electron mass, $\sigma
=\pm 1$ for the electron spin up and down. In a ferromagnetic
metal-superconductor heterostructure, the change of \ the spin direction of
the incident electron (hole) under the Andreev reflection shifts the reflected
hole (electron) into the other energy band that causes an additional
difference $\delta p=I_{0}/v_{F}$. The latter drastically decreases the
penetration length ${\it L}$ by orders of the magnitude \footnote{For $I_{0}\gg kT$
one should change $\Delta p$ in Eq.(\ref{corrdistance}) to $\delta = I_0/v_F$ and
the penetration length inside a diffusive ferromagnet turned out to be  $L_{I_{0}}=\sqrt{\hbar D/I_{0}}$;
the superconducting correlator inside a ballistic ferromagnet oscillates  decreasing linearly
with an increase of the distance from the ferromagnet-superconductor interface \cite{Demler}.}
that influences the quantum interference affecting many properties of
ferromagnet/superconductor heterostructures \cite{Bulaevski,Falko,Fierz,Kawaguchi,Antonov,Jiang,Dong,Shi,Vasko,Upadkhyay,Soulen,Goldman,Leadbeater,Falko2,%
Nazarov,Lazar,Chandrasekhar,Aprili,Oboznov,Zhu,Dubonos,Xia,Krawiec,Radovich1,Radovich2,Radovich3,Burmistrov,Koren,Blanter}.
Such a shortening of the proximity effect has been actually observed in
magnetic materials Refs.~\cite{Kawaguchi,Jiang,Soulen,Goldman,Lazar}. On the
other hand, measurements carried out in recent works \cite
{Petrashov1,Petrashov2,Giordano,Giroud,Giroud1} demonstrate a long-range proximity
effect in magnetic materials that is in an obvious contradiction with the
above general considerations. \ As shown in papers \cite{VolkovEfetov,epl}
such a long-range proximity effect arises in F/S heterostructures in which
the ferromagnetic part is magnetically inhomogeneous. In this case  a triplet superconducting
correlation  persists on a length typical
for {\it non-magnetic }materials \cite{VolkovEfetov,epl,Larkin,Volkov3,Volkov4,Melin,Volkov1,AK,Volkov2,Kopu,Fominov}.
\ A long-range proximity effect of another type occurs in F/S
heterostructures in which the ferromagnetic part is magnetically homogeneous
but the exchange-interaction energy $I_{0}$ is less than the superconductor
energy gap $\Delta$. In this case the difference between the electron and  hole momenta  of Andreev hybrids
 with energy $\varepsilon =I_0$ do  not depend on $I_0$ (see the section \ref{QUALITATIVE}) and hence their
 penetration length is equal  the one  for non-magnetic conductors ${\cal L}_{\varepsilon}$  \cite{sfsdif,Buzdin}
It results
in giant oscillations of the conductance of an S/F/S Andreev interferometer
with a change of the phase difference between the superconductors when the
voltage V applied to the ferromagnet is close to $2I_{0}/e$ \cite{sfsdif}; for the case of F/S structures the subgap conductance
exhibits a peak below the superconducting energy gap $\Delta$ at $e V =I_0 < \Delta$ \cite{Leadbeater}

\ It is important for applications that the both above-mentioned types of
the long-range proximity effects allow to create superconducting quantum
interference devices \ with \ a\ ferromagnetic normal metal junction of an
anomalous large length (a device of such a type was suggested in \cite
{patent}).

In this paper we consider the density of states, the differential conductance and the
current-voltage characteristic (CVC) of an
superconductor-ferromagnet-superconductor heterostructure (S/F/S) of the
Andreev interferometer type in which the ferromagnetic part is
separated from the reservoir of normal electrons with a potential barrier of low
transparency \footnote{The conductance of an S/F/S structure for a diffusive case in the absence of
such a barrier was considered in Ref. \cite{Volkov5}}, $t_{r}\ll 1$, (see Fig.\ref{sample}). For the case that
the exchange-interaction energy $I_{0}$ is less than the superconductor
energy gap $\Delta $ we predict a sharp peak in  the density of states at energies close to $I_0$ and sharp dependence of the differential
conductance $G(V,\varphi )$ on the applied voltage $V$ that is accompanied with
giant oscillations of the density of states and the conductance with a change of the phase difference $%
\varphi $ between the superconductors. For the geometry under consideration the
differential conductance is proportional to the density of Andreev states
that allows their direct spectroscopy with electrical measurements. On the other hand, the predicted effect allows to find
the electron spin - ferromagnet moment exchange interaction energy $I_0$ with a direct  electric measurement because a sharp and high peak  in the differential conductance takes   place exactly  at $V =2 I_0$.

 In section \ref{QUALITATIVE} we  qualitatively explain the phenomenon under consideration. In sections \ref{BDOS} and \ref{BCOND} we present analytical and numerical calculations for the density of states and for the current and the differential conductance in
 ballistic S/F/S structures, respectively. In section \ref{DIFFDOS} we calculate the density of states in a  disordered S/F/S structure
 using the Gutzwiller path-integral approach.
\section{Giant conductance oscillations; qualitative considerations \label{QUALITATIVE}}
\begin{figure}
  \centerline{\includegraphics[width=8.0cm]{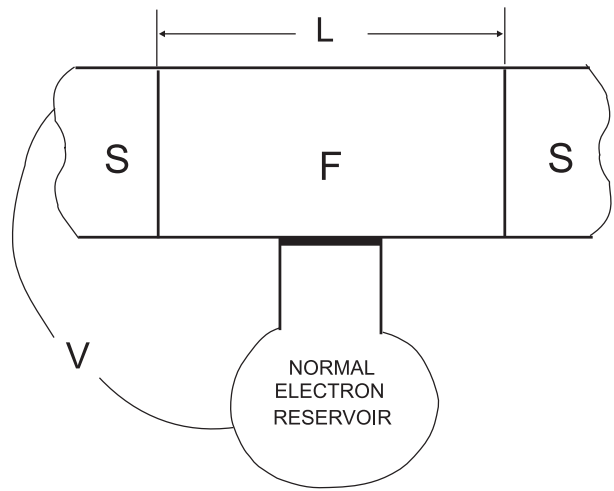}}
  \vspace*{2mm}
  \caption{Superconductor-ferromagnet-superconductor heterostructure of the Andreev
interferometer type. Thick lines indicate potential barriers of low
transparency, $t_r \ll 1$,  separating the ferromagnet from the
reservoir of normal electrons;  voltage $V$ is applied between the reservoir
and the superconductor.}
  \label{sample}
  \end{figure}

In this section we qualitatively show that a change of the superconductor phase difference $\varphi$
can result in giant oscillations of the conductance
of a superconductor-ferromagnetic conductor - superconductor structure if the exchange-interaction
energy $I_{0}$ of the ferromagnet is less than  energy gap $\Delta $ of the superconductor.
The effect arises due to
a resonant transmission of quasi-particles from the
normal reservoir through Andreev states in the ferromagnet which
macroscopically concentrate near the exchange-interaction energy $I_{0}$ if
the superconductor phase difference $\varphi $ is close to an odd number of $%
\pi $ and the voltage $V$ applied to structure (see Fig.\ref{sample}) is close
to $2I_{0}/e$ ($I_{0}<|\Delta |$). A qualitative explanation of the effect
is as follows.

Under Andreev reflection at an F/S interface the spin directions of the
incident and reflected quasi-particles are opposite and hence  the longitudinal momenta (parallel to the $x$-axis which is
perpendicular to the F/S interfaces) of the electron and the hole are,
respectively, (de Jong and Beenakker \cite{Beenakker2})
\[
p_{{\bf n,}\sigma }^{(e)}(\varepsilon )=\sqrt{p_{F}^{2}-p_{\perp }^{2}({\bf n%
})+2m(\varepsilon -\sigma I_{0}})
\]
\begin{equation}
p_{{\bf n,}\sigma }^{(h)}(\varepsilon )=\sqrt{p_{F}^{2}-p_{\perp }^{2}({\bf n%
})-2m(\varepsilon -\sigma I_{0}})  \label{momenta}
\end{equation}
where $p_{F}$ is the Fermi momentum, 
$p_{\perp }({\bf n})=(0,\hbar n_{y}/d_{y},\;\hbar n_{z}/d_{z})$ is the
quantized transverse quasi-particle momentum parallel to the F/S interfaces,
($d_y$ and $d_z$ are the transverse sizes of the ferromagnetic section,
${\bf n}=(0,n_{y},n_{z});\;n_{y},\;n_{z}=0,1,2,...,$ $|{\bf n|\leq }%
N_{\perp }$ are the transverse mode quantum numbers) assuming a hard wall
confining potential with the number of transverse modes $N_{\perp }\approx
S/\lambda _{F}^{2}$ inside it, $S=d_y d_z$ is the  cross section area of the ferromagnet (equal
to the area of the F/S interfaces), $\lambda _{F}=2\pi \hbar /p_{F}$ is the
Fermi wave length.
From Eq. (\ref{momenta}) it follows that in contrast to the non-magnetic
case, near the Fermi level ($\varepsilon \approx 0$) the electron and the
hole momenta in the ferromagnet are different, and for a large enough $I_{0}$
(usually $I_{0}$ is greater than the Thouless energy ) the interference
effects are absent due to the destructive interference. This fact
demonstrates the conflict between superconductivity and magnetic ordering in
S/F/S structures.

However, interference effects in the ferromagnet can exist albeit at some
finite voltage $V$ applied between the reservoir and the superconductor. Our argument starts from
a description of the electron transport in terms of resonant tunneling
through quantized energy levels of the ferromagnet mesoscopic part of the system
shown in Fig. \ref{sample}.

Taking into account the amplitude of the Andreev reflection at a  normal
conductor - superconductor interface \cite{Andreev}
\begin{equation}
r_A = \exp{(i \pi/2 + \phi)}
\label{Areflection}
\end{equation}
one easily finds the semiclassical quantization condition for an S/F/S system in the absence of potential barriers
at F/S interfaces to be \cite{Kulik}
\begin{equation}
\left( p_{{\bf n,}\sigma }^{(e)}(\varepsilon )-p_{{\bf n,}\sigma
}^{(h)}(\varepsilon )\right) L/\hbar +\pi \pm \varphi =2\pi l  \label{dispeq}
\end{equation}
where $p_{{\bf n,}\sigma }^{(e)}(\varepsilon )$ and $p_{{\bf n,}\sigma
}^{(h)}(\varepsilon )$ are the longitudinal momenta determined by Eq. (\ref
{momenta}); $L$ is the distance between the F/S interfaces; $\varphi = \phi_1 - \phi_2$  is the
phase difference between the superconductors 1 and 2; $l=0,\pm
1,\pm 2,..$ is the longitudinal quantum number;
while writing
equations Eq.(\ref{Areflection}) and Eq.(\ref{dispeq}) we assumed
$\varepsilon \ll |\Delta |$ for simplicity's sake.
Dispersion equation Eq.(\ref{dispeq}) determines Andreev levels $\varepsilon_{l,{\bf n}}$
which can be shifted with a change of $\varphi $ by means, for instance, of an
external magnetic field. When an Andreev level is lined with the energy of electrons
 injected from the reservoir of normal electrons, the resonant transmission
 of electrons through the ferromagnetic section takes place that causes an increase of the system
 conductance.
Typically each Andreev level is only 2-fold (spin up and down) degenerated,
and hence simultaneously
only 2 electrons  can  be  resonantly transmitted
through this energy level. It means
that  the amplitude of  oscillations of the
differential conductance $G(V)=dJ/dV$ ($J$ is the dissipative current
current, see Fig. \ref{sample}) with a change of the superconductor phase difference
$\varphi$ is of order of $e^{2}/h$. The situation drastically changes  when the superconductor
phase difference is equal to an odd number of $\pi $. In this case  the energy level $
\varepsilon =I_{0}$ is highly degenerated since all $N_{\perp }$
transverse modes are  simultaneously at this level  (if $\varphi =\pi (2l+1)$ the
dispersion equation Eq.(\ref{dispeq}) is satisfied with $\varepsilon =I_{0}$
for all transverse modes because $p^{(e)}=p^{(h)}$ for any ${\bf n}$, see
Eq.(\ref{momenta}) and Eq.(\ref{dispeq})). Therefore the resonant transmission occurs
simultaneously in all transverse modes when the superconductor phase
difference is equal to odd numbers of $\pi $ and the applied voltage takes
such a value that  $eV/2$ is equal to the exchange energy $I_{0}$, thus
producing a giant conductance peak. The width of the peak in the dependence of the
conductance on $\varphi$ at $V=2I_{0}/e$
is of the order of
the single-electron transparency $t_{r}$ of the potential barriers between
the ferromagnet and the electron reservoir, and the width of the peak in the $V$-dependence
of the conductance at $\varphi =\pi (2l+1)$
is $\sim t_{r}E_{0}$ ($E_{0}=\hbar v_{F}/L$  is the distance between neighboring energy levels); the
amplitude of the peak is of the order of  $N_{\perp }e^{2}/h$ reflecting the
above-mentioned $N_{\perp }$-fold degeneracy of the resonant level. When
this result is compared with that for a non-magnetic normal part (see \cite
{GO,Blom}) it is apparent that the exchange interaction of the electron spin
with the ferromagnet spontaneous momenta destroys the giant oscillations of
the conductance on the Fermi level (that is at $V$ close to zero) shifting
them to the energy range $eV/2\approx I_{0}$ where the giant conductance
oscillations are restored even if $I_{0}\geq E_{0}$.

Giant conductance oscillations for a diffusive ferromagnet in the case that
the temperature $T$ satisfies the inequality $t_{r}E_{Th}^{(D)}\ll kT\ll
E_{Th}^{(D)}=\hbar D/L^{2}$ were considered in paper \cite{sfsdif}. In section\ref{DIFFDOS}
we show that the high degeneracy of the energy level $\varepsilon =I_0$  at $\varphi =\pi$ is not lifted in diffusive
S/F/S structures that results in a sharp peak in the density of states and hence in giant conductance oscillations

\section{Density of states in ballistic S/F/S structures with $I_0 < |\Delta|$ \label{BDOS}}

 In this Section we find the dispersion equation and the density of Anreev states in a
 ferromagnetic conductor-superconductor- ferromagnetic conductor structure in which
potential barriers are present at the ferromagnet-superconductor interfaces.

In order to find  Andreev energy levels $\varepsilon _{{\bf n,}
l}^{\sigma }$ inside the ferromagnetic conductor we use the Stoner model with the exchange
interaction energy $I_{0}$ so that the Bogolubov-de Gennes equations are
written as follows \cite{Beenakker2}:
\begin{eqnarray}
\left \{
\begin{array}{l}
\left( \hat{H}_{0}+\sigma I({\bf r})-\varepsilon \right) u_{+\sigma }+\Delta
v_{-\sigma }=0\\
\Delta ^{\ast }u_{+\sigma }-(\hat{H}_{0}-\sigma I({\bf r})+\varepsilon
)v_{-\sigma }=0
\end{array} \right.
\label{Bogolubov}
\end{eqnarray}
where $\hat{H}_{0}=\hat{{\bf p}}^{2}/2m-\epsilon _{F}$; the superconducting
energy gap $\Delta ({\bf r})$ and the ferromagnet exchange energy\ $I({\bf r}%
)$ have non-zero values in complementary space regions, $\Delta =const\neq 0,
$ $I=0$ in the superconductor and $\Delta =0$, $I=I_0 =const\neq 0$ in the
ferromagnet. We assume Andreev reflections of quasi-particles at the F/S
interfaces to be accompanied by  normal reflections, that is the
scattering process at an F/S interface is described by a $2\times 2$\
scattering matrix \cite{Shelankov,Blonder} as follows \footnote{As was shown by M.J.M. de Jong and C.W.J. Beenakker \cite{Beenakker2}
the scattering matrix at an F/S interface differs from the one  at an N/F interface, but estimations show Eq.(\ref{snmatrix},\ref{phase}) to be correct to the first order  of $I_0 /\varepsilon_F \ll 1$}:
\begin{eqnarray}
\widehat{T}= & e^{i\chi}\left(
\begin{array}{ll}
r_{N}\exp (i\eta ) & r_{A} \\
-r_{A}^{\ast} & r_{N}^{\ast }\exp (-i\eta )\\
\end{array}
\right)
\label{snmatrix}
\end{eqnarray}
where
\begin{eqnarray}
e^{i\chi }=-i\frac{\sqrt{|t_{0}|^{4}+4|r_{0}|^{2}\sin ^{2}\psi _{\varepsilon
}}}{\exp (-i\psi _{\varepsilon })-|r_{0}|^{2}\exp (i\psi _{\varepsilon })};\
\nonumber\\ e^{\psi_\varepsilon }=\frac{|\Delta |}{\varepsilon +i\sqrt{|\Delta
|^{2}-\varepsilon }}\ ;\ \hspace{1.9cm}
\label{phase}
\end{eqnarray}
In Eq.(\ref{snmatrix}) the probability amplitudes of Andreev $(r_{A})$ and \
normal $(r_{N})$\ reflections are written as follows: \
\[
r_{A}=\frac{i|t_{0}|^{2}\exp (i\phi )}{\sqrt{|t_{0}|^{4}+4|r_{0}|^{2}\sin
^{2}\psi _{\varepsilon }}};\ \ \ \ r_{N}=\frac{2r_{0}\sin \psi _{\varepsilon
}}{\sqrt{|t_{0}|^{4}+4|r_{0}|^{2}\sin ^{2}\psi _{\varepsilon }}};
\]
where $\phi $ is the phase of the superconductor energy gap $\Delta
=|\Delta |\exp (i\phi );$ \ $r_{0}$ and $t_{0}$ are the probability
amplitudes for an incident electron to be reflected back and to be
transmitted through the interface in case that the conductors on the both
sides of the interface are in the normal state; this scattering arises due
to a potential barrier at the interface, mismatch between the Fermi
velocities of electrons of the conductors or of \ their effective masses,
and so on. In general case this scattering is described \ by \ the
scattering matrix
\[
\widehat{\rho }=e^{i\eta }\left(
\begin{tabular}{ll}
$|\ r_{0}|$ $exp(i\vartheta )$ & $i|t_{0}|$ \\
$i|t_{0}|$ & $|\ r_{0}|$ $exp(-i\vartheta )$%
\end{tabular}
\right) ;\ \ \ |r_{0}|^{2}+|t_{0}|^{2}=1\ \
\]
According to Eq.(\ref{Bogolubov}) the electron- and hole-like components of
the wave-function in the ${\bf n}$-th transverse mode inside the ferromagnet is

\bigskip
\begin{eqnarray}
u_{\alpha }(x,y)=\sum_{\{{\bf n}\}}\left(
a_{{\bf n}}^{(e)}e^{ik_{{\bf n}}^{(e)}x}+b_{{\bf n}}^{(e)}e^{-ik_{{\bf n}}^{(e)}x}\right)
\sin{k_{\perp }({\bf n})y} \nonumber\\
v_{\alpha }(x,y)=\sum_{\{{\bf n}\}}\left(
a_{{\bf n}}^{(h)}e^{-ik_{{\bf n}}^{(h)}x}+b_{{\bf n}}^{(h)}e^{ik_{{\bf n}}^{(h)}x}\right) \sin
k_{\perp }({\bf n})y%
\label{wf}
\end{eqnarray}
Here $a_{{\bf n}}^{(e)}$ and $b_{{\bf n}}^{(e)}$ [$a_{{\bf n}}^{(h)}$ and $b_{{\bf n}}^{(h)}$] are
the probability amplitudes for free motion of electrons [holes] forward and
backward, respectively, in channel ${\bf n}$; $\ k_{{\bf n}}^{(e,h)}=p_{{\bf n,}%
\sigma }^{(e)}/\hbar $ (see Eq.(\ref{momenta})) is the electron (hole)
longitudinal momentum for an electron (hole) in the ${\bf n}$-th transverse
mode and its spin direction $\sigma $ ($\sigma =\pm 1$), while $x$ and ($%
y,z) $ are longitudinal and transverse coordinates in the sample,
respectively; here and below $\{{\bf n}\}$ stands for the summation  over  all the open modes $|{\bf n}|\leq N_\perp$.

Matching the wave-functions Eq.(\ref{wf}) at two nonequivalent F/S boundaries
with the use of the scattering matrix Eq.(\ref{snmatrix}) results in a spectral
function of the form

\begin{equation}
D_{{\bf n}}^{(\sigma )}(\varepsilon )=\cos \varphi
_{-}-|r_{N}^{(1)}||r_{N}^{(2)}|\cos \varphi
_{+}+|r_{A}^{(1)}||r_{A}^{(2)}|\cos \varphi  \label{dsfun}
\end{equation}
Here $\varphi _{-}(\varepsilon)=(k_{{\bf n}}^{(e)}-k_{{\bf n}}^{(h)})L+\chi _{+}$ and $%
\varphi _{+}(\varepsilon)=(k_{{\bf n}}^{(e)}+k_{{\bf n}}^{(h)})L+\mu _{+}$ where $L$ is
the length of the ferromagnet, $\chi _{+}(\varepsilon )=\chi _{1}+\chi _{2};$
$\mu _{+}=\eta _{1}+\vartheta _{1}+\eta _{2}+\vartheta _{2},$ and $\varphi
=\phi _{1}-\phi _{2}$ is the phase difference between the
superconductors; the indices $1$ and $2$ indicate the F/S boundaries. The
Andreev discrete energy levels $\varepsilon _{{\bf n,}l}^{\sigma }$\ of the
system are determined by solutions of the equation $D_{{\bf n}}^{(\sigma
)}(\varepsilon )=0$, that is
\[
\cos \left( (k_{{\bf n,\sigma }}^{(e)}-k_{{\bf n,\sigma }}^{(h)})L+\chi
_{+}\right) \]
\begin{equation}
|r_{N}^{(1)}||r_{N}^{(2)}|\cos \left( (k_{{\bf n,\sigma }}^{(e)}+k_{{\bf %
n,\sigma }}^{(h)})L+\mu _{+}\right) -|r_{A}^{(1)}||r_{A}^{(2)}|\cos \varphi
\label{dspeq}
\end{equation}

In the case that the ferromagnetic part of the structure is coupled to a reservoir of normal electrons
through a potential barrier of a low transparency $t_r \ll 1$ (see Fig. \ref{sample}), the density of Andreev  states inside the ferromagnetic conductor can be written as follows:
\begin{equation}
\nu (\varepsilon )=\frac{1}{{\cal V}}\sum_{\sigma=-1}^{+1}\sum_{\{{\bf n}\}}\overline{\nu} _{{\bf n}}^{\sigma }(\varepsilon )
\label{fullDS}
\end{equation}
where ${\cal V}$ is the volume of the ferromagnetic part, $\overline{\nu} _{{\bf n}}^{\sigma }$ is the density of Andreev states at a fixed  transverse mode quantum number ${\bf n}$ and
the spin projection $\sigma$:
\begin{equation}
\overline{\nu}^{\sigma}_{\bf n} (\varepsilon )=\frac{1}{\pi }\sum_l \frac{t_r E_0}{(\varepsilon -\varepsilon _{{\bf n},l}^{\sigma })^2+(t_r E_0)^2}=\frac{1}{\pi }\int_{-\infty}^{+\infty}\frac{t_r E_0}{(\varepsilon -\varepsilon^\prime)^2+(t_r E_0)^2}
\nu^{\sigma}_{\bf n}(\varepsilon^\prime)d\varepsilon^\prime
\label{SD}
\end{equation}
and
\begin{equation}
\nu^{\sigma}_{\bf n}(\varepsilon)=\sum_{l}\delta (\varepsilon
-\varepsilon _{{\bf n},l}^{\sigma })
\label{broad}
\end{equation}

For convenience sake of further analytical calculations we write the density
of Andreev states  in the following form $\ \ \ $ \ \ \

\begin{equation}
\nu _{{\bf n}}^{\sigma }(\varepsilon )=\sum_{l}\delta (\varepsilon
-\varepsilon _{{\bf n},l}^{\sigma })\ =\left| \frac{\partial D_{{\bf n}%
}^{(\sigma )}(\varepsilon )}{\partial \varepsilon }\right| \delta \left( D_{%
{\bf n}}^{(\sigma )}(\varepsilon )\right)  \label{density}
\end{equation}
where $D_{{\bf n}}^{(\sigma )}(\varepsilon )$ is defined by Eq.(\ref{dsfun}).

To find the density function $\nu _{{\bf n}}^{\sigma }(\varepsilon )$ we use
the method developed in Ref. \cite{Slutskin}. As $\partial \varphi
_{-}/\partial \varepsilon \approx \pm 1/E_{{\bf n}}$ (here $E_{{\bf n}}=\hbar v_{{\bf n}}/L$ and $v_{{\bf n}}=\sqrt{%
p_{F}^{2}-p_{\perp }^{2}({\bf n})}/m$)  the factor $\partial D_{
{\bf n}}^{(\sigma )}(\varepsilon )/\partial \varepsilon $ is a
trigonometrical function of $\varphi _{\pm }$, as well as $D_{{\bf n}}^{(\sigma )}$ (see Eq.(\ref{dsfun})),
and it is productive to expand $\nu _{{\bf n}}^{\sigma }$ into Fourier
series in $\varphi _{\pm }$ and write it as follows:
\begin{equation}
\nu _{{\bf n}}^{\sigma }(\varepsilon )=\sum_{k=-\infty }^{\infty
}\sum_{s=-\infty }^{\infty }A_{s,k}^{{\bf n},\sigma }\exp (is\varphi
_{-}(\varepsilon)+ik\varphi _{+}(\varepsilon));\   \label{stdensity1}
\end{equation}
\begin{equation}
A_{s,k}^{{\bf n},\sigma }=\frac{1}{(2\pi )^{2}}\int_{0}^{2\pi }d\varphi
_{+}\int_{0}^{2\pi }d\varphi _{-}\left| \frac{\partial D_{{\bf n}}^{(\sigma
)}(\varepsilon )}{\partial \varepsilon }\right| \delta \left( D_{{\bf n}%
}^{(\sigma )}(\varphi _{+},\varphi _{-})\right) \exp (-is\varphi
_{-}-ik\varphi _{+})  \label{A}
\end{equation}
As shown in Ref.\cite{Blom} the main contribution to the state density
function Eq.(\ref{stdensity1}) is of the terms $A_{s,0}^{{\bf n},\sigma }$ which,
after integrating Eq.(\ref{A0}) with respect to $\varphi _{-}$, is written
as \
\begin{equation}
A_{s,0}^{{\bf n},\sigma }=\left( \frac{2}{E_{{\bf n}}}+\frac{\partial \chi
_{+}}{\partial \varepsilon }\right) \frac{1}{(2\pi )^{2}}\int_{0}^{\pi
}d\varphi _{+}\left\{ \exp \left[ is\varphi _{1}(\varphi _{+})\right] +\exp %
\left[ -is\varphi _{1}(\varphi _{+})\right] \right\}  \label{A0}
\end{equation}
where $$\varphi _{1}(\varphi _{+})=\arccos (|r_{N}^{(1)}||r_{N}^{(2)}|\cos
\varphi _{+}-|r_{A}^{(1)}||r_{A}^{(2)}|\cos \varphi ),\hspace{0.5cm} 0\leq \varphi_1\leq \pi$$
Inserting Eq.(\ref{A0}) into Eq.(\ref{stdensity1}) (other terms in Eq.(\ref{stdensity1}) with $%
k\neq 0$ are negligibly small, see Ref.\cite{Blom}) one find $\nu _{{\bf n}}^{\sigma }$ to be
\begin{equation}
\nu _{{\bf n}}^{\sigma }(\varepsilon )=\left( \frac{2}{E_{{\bf n}}}+\frac{\partial \chi
_{+}}{\partial \varepsilon }\right) \sum_{s=-\infty }^{\infty} \frac{1}{(2\pi )^{2}}\int_{0}^{\pi
}d\varphi^\prime\left\{ \exp \left[ is\left(\varphi _{1}(\varphi^\prime )+\varphi
_{-}(\varepsilon)\right)\right] +\exp%
\left[ is\left(-\varphi _{1}(\varphi^\prime)+\varphi
_{-}(\varepsilon)\right)\right] \right\}
\label{DS3}
\end{equation}

Using Eq.(\ref{DS3}), Eq.(\ref{SD}) and Eq.(\ref{fullDS}) one finds the
the density of electronic states in the ferromagnet  to be equal to
\[
\nu(\varepsilon )=\frac{4N_{\perp }}{\pi E_0}\sum_{\sigma =-1}^{1}\sum_{\sigma ^{\prime
}=-1}^{1}\int_{1}^{\infty }\frac{dy}{y^{2}}\left( 2+\frac{E_{0}}{y}\frac{%
d\chi _{+}(\varepsilon )}{d\varepsilon }\right) \times
\]

\begin{equation}
\int_{0}^{\pi }\frac{d\varphi _{+}}{2\pi }\frac{t_{r}}{1-\cos \left[
2(\varepsilon +\sigma I_{0})y/E_{0}+\chi _{+}(\varepsilon )+\sigma ^{\prime
}\varphi _{1}(\varphi _{+})\right] +2t_{r}^{2}}
\label{dens}
\end{equation}

\begin{figure}
 \centerline{\includegraphics[width=14.0cm]{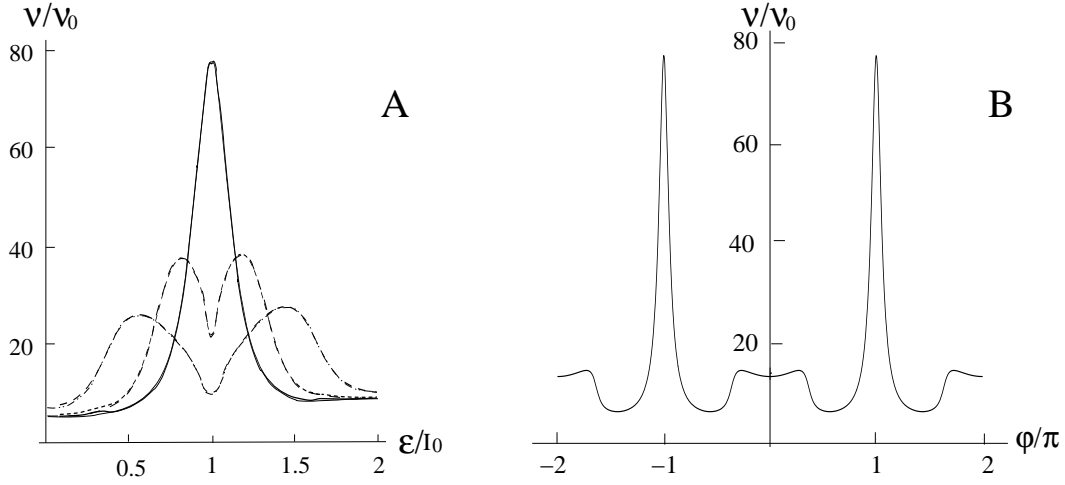}}
  \vspace*{2mm}
  \caption{Normalized density of electronic states inside the ferromagnetic conductor in a ballistic S/F/S structure
for $I_0 = 0.5 E_{0}$, and  $r_N^{(1)}=0.1$, $r_N^{(2)}=0.05$ and $t_r =0.1$;  $\nu_0=8 N_\perp /(\pi E_0) $;. Fig. A shows the dependence of the density of states on
energy $\varepsilon$
  at $\varphi = \pi$
(full line), $\varphi =1.1\pi$ (dash line), and $\varphi =1.2\pi$ (dash-dashed line); Fig. B
shows giant oscillations of the density of states with a change of the superconductor phase difference $\varphi$ at $\varepsilon=I_0$.}
\label{ds}
  \end{figure}

Numerical results for the density of states  based on
Eq.(\ref{dens}) are shown in Fig.\ref{ds}. The sharp  peak in the dependence of the density of states
on energy
corresponds to the $N_\perp$-fold degeneracy of the energy level $\varepsilon =I_0$
in the case that the superconductor phase difference $\varphi$ is equal to odd numbers of $\pi$, as was
explained in the previous section.

\section{NORMAL CURRENT AND CONDUCTANCE OF  S/F/S STRUCTURES with $I_0<|\Delta|$ \label{BCOND}}

In this Section we find the differential conductance and the current through
the ferromagnet in the geometry of Fig. \ref{sample} \ for the case that the ballistic
ferromagnet is weakly coupled to a normal electron reservoir (that is $\
t_{r}\ll 1$) with a voltage drop $V$ \ to be applied between the reservoir
and the superconductor. We assume that there are Andreev and normal
reflections at F/S interfaces 1 and 2 with the probability amplitudes $%
r_{A}^{(1,2)}$ and $r_{N}^{(1,2)}$, respectively ($%
|r_{A}^{(1,2)}|^{2}+|r_{N}^{(1,2)}|^{2}=1$).

According to the Landauer-Lambert formula \cite{Lambert} \ the current from
the normal electron reservoir into the superconductor is written as follows:
\begin{eqnarray}
J=\frac{e}{h}\int_{-\infty }^{\infty }\left( f_{0}(\varepsilon -\frac{eV}{2}%
)-f_{0}(\varepsilon +\frac{eV}{2})\right) R_{A}(\varepsilon ); \label{curr1}\\
R_{A}=\sum_{\{{\bf n}\}}\rho _{A}^{({\bf n)}}
\nonumber
\end{eqnarray}
where $f_{0}(\varepsilon )$ is the Fermi distribution function, \ $\rho
_{A}^{({\bf n)}}$ is the probability for an electron approaching the
ferromagnet along the lead to be reflected back into the reservoir as a hole.

As we are interested in the resonant transmission of quasi-particles through
Andreev levels $\varepsilon _{{\bf n,}l}^{\sigma }$ inside the ferromagnet
we use the resonant Breight-Wigner formula to write $R_{A}(\varepsilon )$ in
the form
\begin{equation}
R_{A}(\varepsilon )=\frac{1}{\pi}\sum_{\sigma =-1}^{1}\sum_{\{{\bf n}\}}
\sum_{l}\frac{t_{r}^{2}}{[(\varepsilon -\varepsilon _{{\bf n,}l}^{\sigma
})/E_{{\bf n}}]^{2}+t_{r}^{2}}  \label{probability}
\end{equation}
 For convenience sake we represent
Eq.(\ref{probability}) as follows:
\begin{equation}
R_{A}(\varepsilon )=\frac{1}{\pi}\sum_{\sigma =-1}^{+1}\sum_{\{{\bf n}\}}\int
\frac{t_{r}^{2}}{[(\varepsilon -\varepsilon ^{\prime })/E_{{\bf n}%
}]^{2}+t_{r}^{2}}\nu _{{\bf n}}^{\sigma }(\varepsilon ^{\prime
})d\varepsilon ^{\prime };  \label{probability1}
\end{equation}

\ \ \ Inserting \ \ Eq.(\ref{DS3}) \ into Eq.(\ref{probability1})
  one finds the
electron-hole transmission coefficient $R_{A}(\varepsilon )$ to be equal to
\[
R_{A}(\varepsilon )=\frac{2N_{\perp }}{\pi}\sum_{\sigma =-1}^{1}\sum_{\sigma ^{\prime
}=-1}^{1}\int_{1}^{\infty }\frac{dy}{y^{3}}\left( 2+\frac{E_{0}}{y}\frac{%
d\chi _{+}(\varepsilon )}{d\varepsilon }\right) \times
\]

\begin{equation}
\int_{0}^{\pi }\frac{d\varphi _{+}}{2\pi }\frac{2t_{r}^{2}}{1-\cos \left[
2(\varepsilon +\sigma I_{0})y/E_{0}+\chi _{+}(\varepsilon )+\sigma ^{\prime
}\varphi _{1}(\varphi _{+})\right] +2t_{r}^{2}}  \label{transprob}
\end{equation}
Using Eq.(\ref{transprob}) and Eq.(\ref{curr1}) one finds the differential
conductance $G(V)=dI/dV$ to be
\[
G(E)=\frac{2e^{2}N_{\perp }}{\pi h}2\sum_{\sigma =-1}^{1}\sum_{\sigma ^{\prime
}=-1}^{1}\int_{1}^{\infty }\frac{dy}{y^{3}}\int_{0}^{\pi }\frac{d\varphi _{+}%
}{2\pi }\times
\]
\begin{equation}
\frac{2t_{r}^{2}}{1-\cos \left[ (e\overline{V}+2\sigma \overline{I}%
_{0})y+\sigma ^{\prime }\varphi _{1}(\varphi _{+})\right] +2t_{r}^{2}}
\label{conduct}
\end{equation}
where $\overline{V}=V/E_{0}$ and $\ \overline{I}_{0}=I_{0}/E_{}$, and the
current to be
\[
J=\frac{8eN_{\perp }E_{0}}{\pi h}\frac{t_{r}}{\sqrt{1+t_{r}^{2}}}\sum_{\sigma
=\pm 1}\int_{1}^{\infty }\frac{dy}{y^{4}}\int_{0}^{\pi }\frac{d\varphi _{+}}{%
2\pi }\times
\]

\[
\left\{ \arctan \left[ \frac{\sqrt{1+t_{r}^{2}}}{t_{r}}\tan \left( \frac{%
(eV/2+I_{0})y}{E_{0}}+\sigma \frac{\varphi _{1}(\varphi _{+})}{2}\right) %
\right] -\right.
\]
\begin{equation}
\left. \arctan \left[ \frac{\sqrt{1+t_{r}^{2}}}{t_{r}}\tan \left( \frac{%
(-eV/2+I_{0})y}{E_{0}}+\sigma \frac{\varphi _{1}(\varphi _{+})}{2}\right) %
\right] \right\}  \label{cur}
\end{equation}

\begin{figure}
 \centerline{\includegraphics[width=8.0cm]{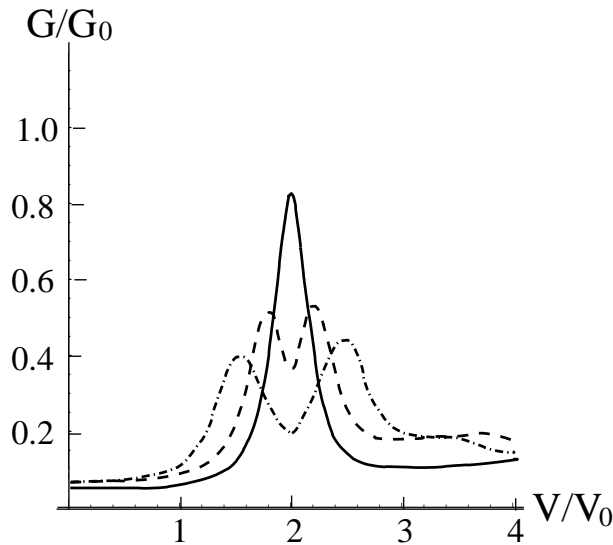}}
  \vspace*{2mm}
  \caption{Normalized differential conductance $G=dI/dV$ of a ballistic S/F/S structure
for $I_0 = E_{0}$, and  $r_N^{(1)}=0.1$, $r_N^{(2)}=0.05$ and $t_r =0.1$
at $\varphi = \pi$
(full line), $\varphi =1.1\pi$ (dashed line), and $\varphi =1.2\pi$ (dash-dotted line); $G_0=2eN_\perp/h$}
\label{cond}
  \end{figure}

\begin{figure}
 \centerline{\includegraphics[width=8.0cm]{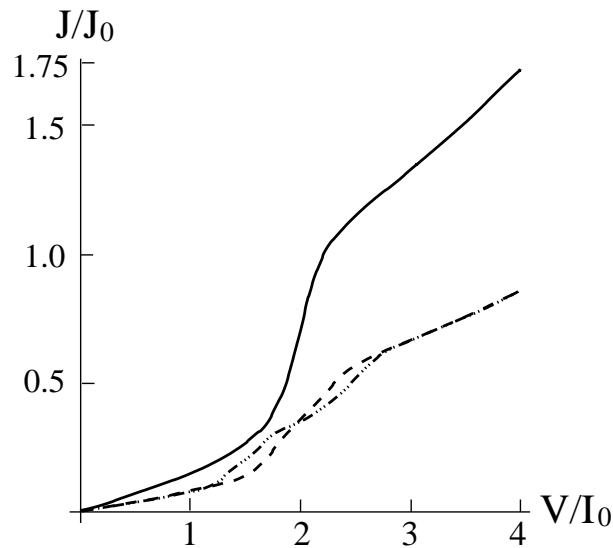}}
  \vspace*{2mm}
  \caption{Normalized current-voltage characteristics for  phase differences
$\varphi = \pi$
(full line), $\varphi =1.1\pi$ (dashed line), and $\varphi =1.2\pi$ (dash-dotted line)
shown for $r_N^{(1)}=0.1$, $r_N^{(2)}=0.05$, $t_r =0.1$, and
$I_0 = E_{0}$; $J_0=2eN_\perp E_0/h$}
\label{zbcur}
  \end{figure}

Numerical results for the differential conductance and the current based on
Eq.(\ref{conduct}) and Eq.(\ref{cur}) are shown in Fig. \ref{cond} and Fig. \ref{zbcur}. They
demonstrate a high sensitivity of the differential conductance and the
non-linear current-voltage characteristics to both the superconductor phase
difference $\varphi $ and the applied voltage $V.$

At low voltages, far from $2I_{0}/e$, we have a resonant tunneling of
quasi-particles through separate Andreev levels, and the conductance and the
current are low. When $eV/2$ $\approx I_{0}$ and $\varphi =\pi (2l+1)$ ($%
l=0,\pm 1,\pm 2,...$) Andreev levels concentrate near $I_{0}$ as can be
readily seen from Eq.(\ref{dsfun}), and we \ have simultaneous resonant
transmission of quasi-particles through the whole number of $N_\perp $ states
resulting in a high peak in the conductance and a sharp jump in the current.
When $\varphi $ deviates from an odd number of $\pi $, the number of Andreev levels concentrated
near $I_{0}$ is decreasing that results in a decrease of the sensitivity of
the conductance and the current to the voltage. \

As is evident from Eq.(\ref{dspeq}),\ in case $r_{N}^{(1)}\neq r_{N}^{(2)}$
\ \ Andreev levels are repelled from the level $I_{0}$.  It results in a splitting
(proportional to $\delta r_{N}=|r_{N}^{(1)}-r_{N}^{(2)}|$ , see \cite{Blom})
of the conductance peak if this splitting is larger than the Andreev level
broadening (proportional to $t_{r}$) caused by the coupling of the
ferromagnet to the reservoir through a potential barrier of the transparency
$t_{r}\ll 1$. In Fig. \ref{cond} this splitting is absent at $\varphi =\pi $ because in
this case $\delta r_{N}=0.05<t_{r}=0.1$. However, with a deviation of the
superconductor phase difference $\varphi $ from $\pi $ the splitting sharply
increases, and a dip in the conductance peak appears as is seen in Fig.\ref{cond} at
$\varphi =1.1\pi $ and $\varphi =1.2\pi .$

We note here that the differential conductance $G$ as a function of $eV$ is
proportional to the the density of Andreev states in the ferromagnet
permitting a direct spectroscopy of Andreev levels by conductance and
current measurements.\ \ \ \

 \section{Density of states  in diffusive S/F/S structures with $I_0 < |\Delta|$  \label{DIFFDOS} }

As is shown in Section \ref{BDOS} for   ballistic S-F-S structures, the energy level $\varepsilon = I_0 < |\Delta|$ is
$N_\perp$-fold degenerated and  the density of states $\nu (\varepsilon)$ at $\varepsilon =I_0$ and $\varphi = \pi$ is proportional to $N_\perp \sim S/\lambda_F^2 \gg 1$ (see Fig. \ref{ds}) in the same manner as it take place in S-N-S (non-magnetic) structures at $\varepsilon = 0$ (see \cite{GO,Blom}). In diffusive non-magnetic S-N-S structures
there is an energy gap in the density of states around
$\varepsilon = 0$ which is maximal at $\varphi = 0$ and shrinking to zero together with a sharp increase of the density of states  at
$\varepsilon = 0$ as $\varphi$ approaches $\pi$ \cite{Spivak}. In order to  see whether the above-mentioned degeneracy
of the level $\varepsilon =I_0 < |\Delta$  survives in diffusive S/F/S structures, we consider
 the density of states in a diffusive S/F/S structure  in the vicinity of  $\varphi =\pi (2k+1)$, $k= 0,\pm 1, \pm 2, ...$
assuming the motion of quasiparticles inside the ferromagnet to be semiclassical. It allows us to use Gutzwiller's approach \cite{Gutzwiller}
which has the advantage of the clarity of physical presentation. On the other hand, the new class of two-dimensional (2D) magnetic semiconductors with large dielectric constants and small effective masses \cite{Smorchkova} very well satisfies the condition of semiclassical motion
$\alpha = r_s p_F /\hbar \gg 1$ ($r_s$ is the screening length in the ferromagnet) with $\alpha = 10 \div 10^2$. We show, that in  much the same way as in the ballistic case (see Section \ref{QUALITATIVE} and Section \ref{BDOS}), in a diffusive S/F/S  structure  the  energy level $\varepsilon = I_0$  ($I_0 < |\Delta|$) is
also  $N_0 \sim S/\lambda_F^2$ -fold degenerated and
the hight of the peak in the density of states at $\varepsilon = I_0$ is proportional to  $N_0$, if the phase difference between the superconductors is in the vicinity of an  odd number of $\pi$.
Below  we consider the case that the Andreev reflection of a quasiparticle at the  F/S interfaces is accompanied by a normal specular reflection in the same manner as for the ballistic case, that is the probability amplitudes $|r^{(1,2)}_N| \ll 1 $.

For calculations of the density of states  $\nu_{SFS}(\varepsilon)$  at $\varphi = \pi(2k+1)$, $k=0, \pm 1, ...$ we use the Gutzwiller semiclassical trace formula
 which  shows  that  $\nu_{SFS}(\varepsilon)$ can be presented as a sum over all periodic classical trajectories \cite{Gutzwiller}:
\begin{equation}
\nu_{SFS} (\varepsilon)= \sum_j A_j(\varepsilon)e^{i S_j (\varepsilon)}
 \label{DDOS}
 \end{equation}
(here $j$ is the trajectory index, $S_j = \oint p dq$ is the classical action integral (the integral is taken along the periodic $j$-trajectory),
$A_j$ is the trajectory amplitude which is equal to the Gaussian path integral around the stationary phase path (which  is the classical trajectory)).

\begin{figure}
 \centerline{\includegraphics[width=14.0cm]{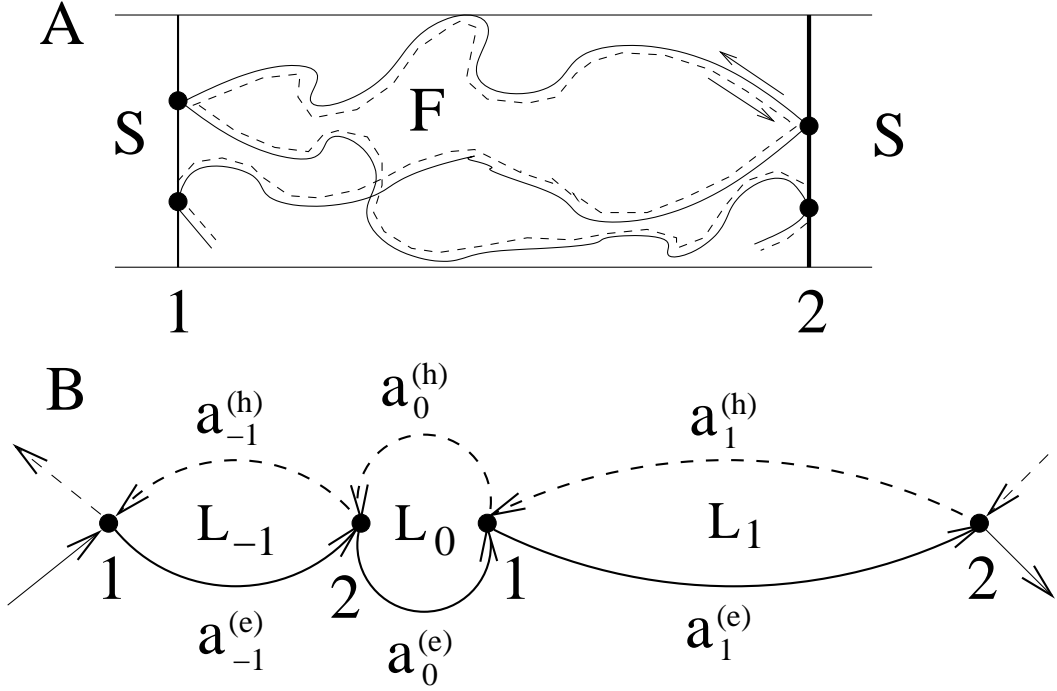}}
  \vspace*{2mm}
\caption{A, Structure analyzed in the text with superconducting (S) and disordered ferromagnetic (F) elements  labeled. The ferromagnetic element is coupled to two superconductors. Semiclassical trajectories of electron- and hole-like guasiparticle excitations are indicated by solid and dashed lines, respectively. Sections of the trajectories going between F/S boundaries  1 and 2 are connected by Andreev and normal reflections at the points shown as black dots. B, Sequence of scattering events along the trajectory shown in A; $L_n$ is the length of the trajectory in section $n$.}
\label{path}
 \end{figure}

If there are Andreev and  normal reflections at the F/S boundaries,  the semiclassical trajectories contributing to  the density of states Eq.(\ref{DDOS}) are of the type shown in Fig. \ref{path}.
In this case the problem of finding the density of states is reduced to a quantum scattering problem for the configuration shown in Fig. \ref{path}B. The points of reflections at the F/S interfaces are shown with black dots.
Propagation between these points is coherent in both electron and hole channels, which is illustrated  by dashed and solid lines of equal lengths. For a quasiparticle with a non-zero energy $\varepsilon$ the phase gains along  electron and hole trajectories do not completely cancel since the momenta are now different. The resulting decompensation effect is of order of $||\varepsilon| -I_0|/E_{Th}$ \footnote{The semiclassical trajectories for an electron of energy $\varepsilon $ and the reflected hole of energy $-\varepsilon$ are to be considered identical since they separate by less than $\lambda_F$ while diffusing a distance $L$ if $||\varepsilon | -I_0| \ll E_{Th}$.}

If $r_N^{(1,2)} \neq 0$, the motion of a quasiparticle along such a trajectory is of a quantum character  despite  the quasiclassical parameter is small because quasi-particle waves split at the scattering points. The semiclassical  wave-functions in the adjacent sections of Fig. \ref{path}B are connected by the 2-channel scattering matrix \ref{snmatrix} describing Andreev and normal reflections at the F/S boundaries. The problem of finding the spectrum and the wave-function in such a one-dimensional chain is reduced to solving a set of matching equations for amplitudes of electron $a_n^{(e)}$
and hole $a_n^{(h)}$ excitations in every section of propagation between scattering points. If $r_N^{(1,2)}\ \ll 1$, the main contribution
to the density of states Eq.(\ref{DDOS}) at $\varphi =\pi(2k+1)$, $k=0,\pm 1, \pm 2, ...$ comes from trajectories which do not include successive reflections at the same F/S boundary. Taking these observation into account the following set of algebraic matching equations emerge:
\begin{eqnarray}
\begin{array}{l}
a_0^{(e)} = r^{(2)}_N e^{i\Phi_{-1}^{(e)}}a_{-1}^{(e)}  + r^{(2)}_A e^{-i\Phi_0^{(h)}}a_0^{(h)}\\
a_{-1}^{(h)} = -r^{(2)*}_A e^{i\Phi_{-1}^{(e)}}a_{-1}^{(e)}  + r^{(2)*}_N e^{-i\Phi_0^{(h)}}a_0^{(h)}\\
a_{1}^{(e)} = r^{(1)}_N e^{i\Phi_{0}^{(e)}}a_{0}^{(e)}  + r^{(1)}_A e^{-i\Phi_{1}^{(h)}}a_{1}^{(h)}\\
a_{0}^{(h)} = -r^{(1)*}_A e^{i\Phi_0^{(e)}}a_0^{(e)}  + r^{(1)*}_N e^{-i\Phi_{1}^{(h)}}a_{1}^{(h)}\\
a_{-1}^{(e)} = r^{(1)}_N e^{i\Phi_{-2}^{(e)}}a_{-2}^{(e)}  + r^{(1)}_A e^{-i\Phi_{-1}^{(h)}}a_{-1}^{(h)}\\
a_{-2}^{(e)} = -r^{(1)*}_A e^{i\Phi_{-2}^{(e)}}a_{-2}^{(e)}  + r^{(1)*}_N e^{-i\Phi_{-1}^{(h)}}a_{-1}^{(h)}\\
... = ...
\end{array}
\label{set}
\end{eqnarray}
The amount of the phase gain after propagation along the trajectories in section $n$ is $$\Phi_{n}^{(e,h)}= \int_{L_n} p_{\sigma}^{(e,h)} d l/\hbar \approx \Phi_{n}^{(0)} \pm  \tau_n (\varepsilon -\sigma I_0)/\hbar$$ where $\Phi_{n}^{(0)}= p_F L_n/\hbar$ and $\tau_n = L_n/v_F$  is the propagation time in section $n$. The quantity $r_A^{(1,2)}$ and $r_N^{(1,2)}$ are, respectively, the probability amplitudes for Andreev and normal reflections at F/S boundaries 1 and 2. Phases and amplitudes of  electrons or  holes along semiclassical paths are defined in such a way that no phase has been gained at the beginning of a particular electron or hole section $n$. Hence the amplitude is $a_n^{(e,h)}$ at the beginning of the section, and $a_n^{(e,h)} \exp{\pm \Phi_{n}^{(e,h)}}$ at its end. We note that one can show that the large phases $\Phi_{n}^{(0)}$ can be removed from the set of eqns (\ref{set}). This is a manifestation of the fact that the electron and hole phase gains compensate each other at $\varepsilon =\sigma  I_0$.

The set of equations Eq.(\ref{set}) can be presented in a matrix form:
\begin{equation}
| {\bf a}> = {\hat U}| {\bf a}>
\label{matreq}
\end{equation}
where components of the vector $| {\bf a}>$  are the coefficients $a_k^{(e,h)}$   of Eq(\ref{set}); the definition of the matrix ${\hat U}$ is obvious from Eq.(\ref{set}): it is  a unitary matrix written as follows:
\begin{equation}
U_{mk}({\bf \Phi}^{(e,h)}) = V_{mk}\exp{\{i \Phi_k^{(e,h)} \}}
\label{U}
\end{equation}
where the $m$-th row of the matrix ${\hat V}$ has only two non-zero elements which are elements of the matrices ${\hat T}^{(1,2)}$ (see Eq.(\ref{snmatrix})); ${\bf \Phi}^{(e,h)} = (..., \Phi_{-1}^{(e,h)},\Phi_1^{(e,h),\Phi_2^{(e,h)}},...)$.

According to general principles of quantum mechanics \cite{Landau}, the spectrum of the system  is  defined by zeros of the determinant
\begin{equation}
Det\left \Vert {\hat I} - {\hat U}({\bf \Phi}(\varepsilon)) \right \Vert =0
\label{Det}
\end{equation}
where ${\hat I} $ is the unit matrix.

Summation in Eq.(\ref{DDOS}) is over different trajectories and one can think of it as averaging over various distributions of $\tau_n$.
Therefore, the density of states Eq.(\ref{DDOS}) for the S/F/S case under consideration can be re-written as follows:
\begin{equation}
\nu_{SFS} (\varepsilon)\approx N_0 \sum_\sigma \left < \left <\nu_{random}^{\sigma} (\varepsilon) \right > \right >
\label{average}
\end{equation}
where $N_0 = S/\lambda_F^2$ and $<< ...>>$ implies an averaging over $\tau_n$; $\nu_{random}^\sigma (\varepsilon)$ is the density of Andreev states for a fixed spin projection $\sigma$ generated by a given semiclassical trajectory of the type shown in Fig. \ref{path}B.
The distribution of propagation times $\tau_n$ depends on details of the disordered potential in the mesoscopic ferromagnetic  region. These are not known, but it is natural to assume that the propagation times along different sections of the semiclassical trajectories
(see Fig. \ref{path}A) are uncorrelated.

In order to find the averaged density of states $<<\nu_{random}^\sigma (\varepsilon)>>$ we use the  presentation suggested by Slutskin \cite{Slutskin} for the density of states
 of a one-dimensional chain of the type in Fig. \ref{path}B. In the general case it  can be written as follows:
\begin{equation}
\nu (\varepsilon) =\frac{1}{\hbar N} \sum_{n=1}^N \tau_n \left (F_{nn} + F_{nn}^{*} + 1 \right)
\label{F}
\end{equation}
where matrix ${\hat F}$ satisfies the matrix equation ${\hat F} = {\hat F} {\hat U} +{\hat U}$ (in our case ${\hat U}$ is defined by Eq.(\ref{U})) and
\begin{equation}
 F_{nn}  = \sum_{{\bf l}} {\bar A}_n ({\bf l}) \exp{\{i {\bf l}{\bf f}\}}
\label{FF}
\end{equation}
Here ${\l} =\{l_1, ...,l_N\}$ ($N$ is the number of sections in the chain), $l_n$ are integers, either positive or equal to zero, ${\bf l}{\bf f} = \sum_{k=1}^{N} l_k f_k $,
 ${\bf f} =\{f_1 (\varepsilon), ...,f_N (\varepsilon)\}$, $f_n (\varepsilon)$ is the phase gain along section $n$ of the chain,
 Fourier coefficients ${\bar A}_n$ depends on "hopping integrals" $r_N^{(1,2)}$ and are independent of $\tau_k$, $k =1, 2, ...N$;
 summation  is over integer  $l_k$, either positive or equal to zero.
Therefore, the averaged density of states is
\begin{equation}
\left < \left < \nu_{random}^\sigma (\varepsilon)\right > \right > =\frac{1}{\hbar N} \sum_{n=1}^N \left < \left <\tau_n \left (F_{n} + F_{n}^{*} + 1 \right)\right > \right >
\label{avdos}
\end{equation}
where $F_n \equiv F_{nn}$.

 For the case under consideration $f_k = \tau_k (\varepsilon -\sigma  I_0)/\hbar$ and for the convenience sake we  re-write   Eq.(\ref{FF})
 as follows:
\begin{equation}
F_{n} = \sum_{{\bf l}} {\bar A}_n ({\bf l})\prod_k \exp{\{i l_k \tau_k (\varepsilon -\sigma  I_0) /\hbar \} }
\label{Fourier}
\end{equation}
Since   amplitudes ${\bar A}_n$ do not depend on $\tau_k$ one only has to average the product $\tau_n \prod_k \exp{i l_k \tau_k (\varepsilon-\sigma  I_0) /\hbar }$ while calculating the average density of states Eq.(\ref{avdos}).

For the configuration of Fig. \ref{path}B the averaged density of states can be found exactly if one
 chooses a Lorentz form for the distribution function $P(\tau)$,
\begin{equation}
P(\tau) =\frac{1}{\pi} \frac{\gamma}{(\tau - \bar \tau)^2 + \gamma^2},
\label{Lorentz}
\end{equation}

Using the Lorenzian distribution Eq.(\ref{Lorentz}) one finds
the result
\begin{equation}
\left< \left< \tau_k \prod_k \exp{i l_k \tau_k (\varepsilon-\sigma  I_0) /\hbar }\right > \right > = \frac{\gamma}{\pi {\bar \tau}}\int_{-\infty}^{\infty}d \varepsilon_1 \frac{\varepsilon_1}{\left(\varepsilon_1 -(\varepsilon -\sigma  I_0)\right)^2 + (\gamma /{\bar \tau})^2 (\varepsilon -\sigma  I_0)^2 }\left({\bar \tau} \prod_k \exp{i l_k {\bar \tau} \varepsilon_1 /\hbar } \right)
\label{Avsum}
\end{equation}
Inserting  Eq.(\ref{Avsum}) in Eq.(\ref{avdos}) one finds the averaged density of states:
\begin{equation}
\left < \left < \nu_{random}^{\sigma} (\varepsilon)\right > \right > = \frac{2 \gamma |\varepsilon -\sigma  I_0|}{\pi {\bar \tau}}\int_{- \infty}^\infty \frac{\varepsilon_1^{2} \nu_p(\varepsilon_1) d \varepsilon_1}{\varepsilon_1^4 + 4(\gamma /{\bar \tau})^4(\varepsilon -\sigma  I_0)^4}
\label{avdensity}
\end{equation}
where
\begin{equation}
\nu_p (\varepsilon) =\frac{1}{\hbar N} \sum_{n=1}^N {\bar \tau} \left (F_{n}^{(0)} + F_{n}^{(0)*} + 1 \right)
\label{F0}
\end{equation}
and
\begin{equation}
F_n = \sum_{{\bf l}} {\bar A}_n ({\bf l})\prod_k \exp{\{i  l_k {\bar \tau} \varepsilon /\hbar\}}
\label{FF0}
\end{equation}
While writing Eq.(\ref{avdensity}) we took into account the fact that $\nu_p (\varepsilon_1) = \nu_p (-\varepsilon_1)$.

From Eq.(\ref{F0}) and Eq.(\ref{FF0})  one sees  $\nu_p (\varepsilon)$  to be  the  density of states
  for the case that the system in Fig. \ref{path}B is non-magnetic ($I_0=0$) and  periodic with all $\tau_n = {\bar {\tau}}$ (that is all  $L_n = {\bar L}$ where ${\bar L} = <<L_n>> \equiv {\bar \tau}/v_F \sim  (v_F /D) L^2$ where $L$ is the distance between the S/F interfaces).

  For the case of a  periodic chain,   Eq.(\ref{set}) with all $\tau_n = {\bar {\tau}}$ can be easily solved, and  for the phase difference $\varphi$ in the vicinity of an odd number of $\pi$ one gets the dispersion low for the quasiparticle moving along the chain as follows:
  \begin{equation}
\varepsilon_{\pm}  (k) =\pm E_{Th}\sqrt{(\delta \varphi)^2 + |r_N^{(1)}|^2 +|r_N^{(2)}|^2 -2|r_N^{(1)}||r_N^{(2)}|\cos{k} }
  \label{dispersion0}
  \end{equation}
where $E_{Th} = \hbar D/L^2$ is the Thouless energy,  $\delta \varphi = \varphi - l \pi $ ($l$ is an odd number, $|\delta \varphi| \ll 1$),  $k$ is a continuous quantum number - the "quasi-momentum" in the periodic  one-dimensional chain.

Using Eq.(\ref{dispersion0}) one finds  the density of states $\nu_p$ of the periodic chain to be
 \begin{equation}
\nu_p(\varepsilon) = \frac{2 |\varepsilon|}{\pi \sqrt{\left( \varepsilon^2 - \varepsilon_{min}^2\right) \left(  \varepsilon_{max}^2  - \varepsilon^2 \right)}}
  \label{density0}
  \end{equation}
  where the minimal and the maximal energies of the energy band Eq.(\ref{dispersion0}) are
\begin{eqnarray}
\begin{array}{l}
\varepsilon_{min}= E_{Th}\sqrt{(\delta \varphi)^2 + \left(|r_N^{(1)} | -|r_N^{(2)} |\right)^2}\\
\varepsilon_{max}= E_{Th}\sqrt{(\delta \varphi)^2 + \left(|r_N^{(1)} | +|r_N^{(2)} |\right)^2}
\end{array}
\label{minmax}
\end{eqnarray}

Inserting Eq.(\ref{minmax}) in Eq.(\ref{avdensity}) and integrating the resulting expression, we find an  exact formula for
the averaged density of states of the random chain in Fig. \ref{path}:

\begin{equation}
\nu_{random}^\sigma(\varepsilon) = \frac{2 \kappa |\varepsilon -\sigma I_0|}{\pi}\left \{\frac{\sqrt{\left(\varepsilon_{min}^4 +4 (\kappa|\varepsilon -\sigma I_0|)^4 \right)\left(\varepsilon_{max}^4 +4 (\kappa|\varepsilon -\sigma I_0|)^4 \right)}+\varepsilon_{min}^2\varepsilon_{max}^2 - 4(\kappa |\varepsilon -\sigma I_0|)^4}{2 \left(\varepsilon_{min}^4 +4 (\kappa
|\varepsilon -\sigma I_0|)^4 \right)\left(\varepsilon_{max}^4 +4 (\kappa|\varepsilon -\sigma I_0|)^4 \right)}\right \}^{1/2}
  \label{exactdensity}
  \end{equation}
where $\kappa=\gamma/{\bar \tau}$.

Multiplying Eq.(\ref{exactdensity}) by $N_0$ and summing over the spin projections
(see Eq.(\ref{average})) we get the average density of states of a diffusive S-F-S structure as follows:
\begin{equation}
\nu_{SFS}(\varepsilon) = N_0 \sum_\sigma \frac{2 |\varepsilon -\sigma I_0|}{\pi}\left \{\frac{\sqrt{\left(\varepsilon_{min}^4 +4 (\varepsilon -\sigma I_0|)^4 \right)\left(\varepsilon_{max}^4 +4 (\varepsilon -\sigma I_0|)^4 \right)}+\varepsilon_{min}^2\varepsilon_{max}^2 - 4(\varepsilon -\sigma I_0)^4}{2 \left(\varepsilon_{min}^4 +4 (\varepsilon -\sigma I_0|)^4 \right)\left(\varepsilon_{max}^4 +4 (\varepsilon -\sigma I_0|)^4 \right)}\right \}^{1/2}
  \label{finaldensity}
  \end{equation}
(we have used $\gamma = {\bar \tau}$, see Ref. \cite{Blom} )

\begin{figure}
 \centerline{\includegraphics[width=14.0cm]{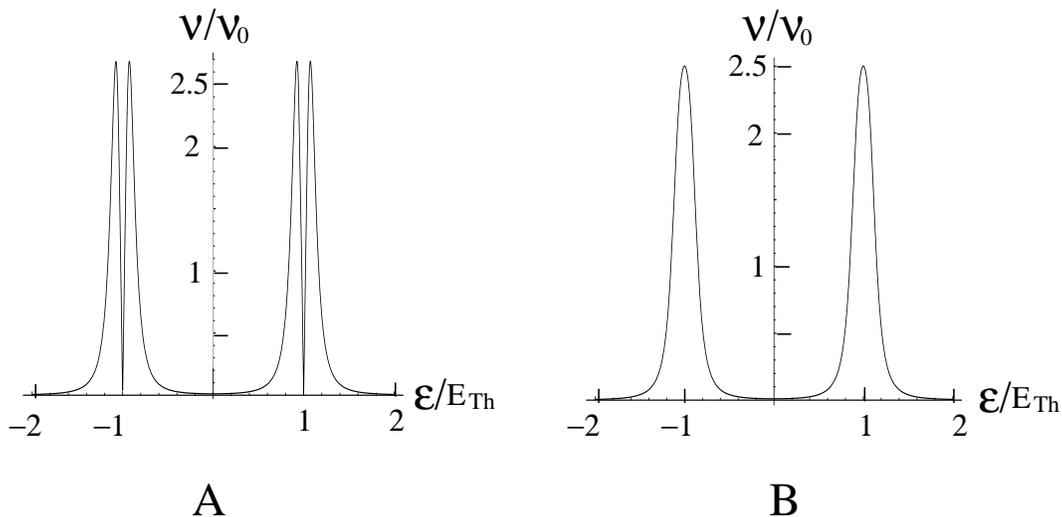}}
  \vspace*{2mm}
  \caption{A, The averaged density of states for a diffusive S/F/S structure for $I_0 = E_{Th}$ and
  $r_n^{(1)}= 0.05$, $r_n^{(2)}= 0.1$,
  $\varphi =\pi$.   B, The averaged density of states for a diffusive S/F/S structure for $I_0 = E_{Th}$ and
  $r_{n}^{(1)}= r_{n}^{(2)}   0.1$; $\nu_0 = 2/(\pi E_{Th})$
  }
\label{sfsdos}
  \end{figure}

  Numerical results for the averaged density of states for a diffusive S/F/S structure based on Eq.(\ref{finaldensity}) are presented in Fig. \ref{sfsdos}.  The sharp peak in the dependence of the averaged density of states on energy In Fig. \ref{sfsdos}B corresponds to the $N_0$-fold degeneracy of the energy level $\varepsilon =I_0$ at $\varphi$ equal to odd numbers of $\pi$; the splitting of the peak in Fig. \ref{sfsdos}A  is proportional to  $||r_N^{(1)}| - |r_N^{(2)}|| $.

\section{Conclusion}

In conclusion, for S/F/S structures with $I_{0}<$\ $\Delta $ ($I_{0}$ is the
ferromagnet exchange energy, $\Delta $ is the superconductor energy gap) we
have demonstrated that an extremely high  degeneracy of the energy level $\varepsilon=I_0$
proportional to  $N_0 = S/\lambda_F^2$  at $\varphi$ equal to odd numbers of $\pi$ ($S$ is the cross-section area of the ferromagnet, $\lambda_F$ is the electron wave length) results in
giant oscillations of the density of Andreev states and the conductance
of the ferromagnet with a change of the superconductor phase difference $\varphi$.
 This
phenomena is a convenient tool for the Andreev level spectroscopy (the
differential conductance $G$ is proportional to the density of Andreev
states) and enables applications, e. g. as a double-gate ferromagnet
transistor analogous to the one described in \cite{patent}. On the other hand, this  effect
 permits  to find the exchange energy, $I_0$, of the interaction between  the electron spin and
the spontaneous moment of the ferromagnetic conductor by  an  electric measurement of the differential
conductance because
a sharp and giant peak in the  dependence of the conductance on the applied voltage $V$ arises
 at $V=2 I_0/e$ (the corresponding peak in the density of Andreev states takes place at energy $\varepsilon =I_0$;
 see Fig.\ref{ds}, Fig. \ref{cond} and Fig. \ref{sfsdos}).

\section*{Acknowledgment}
 A.K. acknowledges the support received from
the European Science Foundation (ESF) 'Novel Applications of
Josephson Junctions in Quantum Digital Circuits (PiSHIFT)'.
 M.J and R.S. acknowledge the financial support from the SSF NANODEV
center and R.S. from the SSF programme on Magnetoelectronic
Nanodevice Physics.

\end{document}